\documentclass[twocolumn,aps,prl]{revtex4-1}

\usepackage{amsmath}
\usepackage{graphicx}
\usepackage{color}
\usepackage{url}
\usepackage[left=0.75in, right=0.75in,top=0.75in,bottom=1.5in]{geometry}

\begin{document}

\title{Strength and scales of itinerant spin fluctuations in 3\emph{d} paramagnetic metals}

\author{Aleksander L. Wysocki}
\email{alexwysocki2@gmail.com}

\author{Andrey Kutepov}

\author{Vladimir P. Antropov}

\affiliation{Ames Laboratory, Ames, IA 50011, USA}

\date{\today}

\begin{abstract}
The full spin density fluctuations (SDF) spectra in 3\emph{d} paramagnetic metals are analyzed from first principles using the linear response technique. Using the calculated complete wavevector and energy dependence of the dynamic spin susceptibility, we obtain the most important, but elusive, characteristic of SDF in solids: on-site spin correlator (SC). We demonstrate that the SDF have a mixed character consisting of interacting collective and single-particle excitations of similar strength spreading continuously over the entire Brillouin zone and a wide energy range up to femtosecond time scales. These excitations cannot be adiabatically separated and their intrinsically multiscale nature should be always taken into account for a proper description of metallic systems. Overall, in all studied systems, despite the lack of local moment, we found a very large SC resulting in an effective fluctuating moment of the order of several Bohr magnetons. 
\end{abstract}

\maketitle

SDF in metals\cite{MoriyaBook,SolontsovBook,TakahashiBook} determine the magnetic dynamics and play an important role in many technologically important applications including Invar alloys\cite{Wassermann}, spintronics\cite{Zhang}, and high temperature superconductors\cite{Moriya} including newly discovered ferropnictides\cite{Dai}. SDF are also crucial in physics of quantum phase transitions including quantum criticallity \cite{QCPReview}. However, a description of SDF in real metallic systems is challenging due to the itinerant nature of the valence electrons. Theories based on the localized Heisenberg model, which are very successful in magnetic insulators, are no longer applicable because of this itineracy. A proper dynamic treatment of essentially quantum SDF is crucial especially at low temperatures. However, the strength and the structure of such itinerant spin fluctuations has not yet been established on neither qualitative nor quantitative level in any material.

The key quantity characterizing SDF in itinerant systems is SC which represents the on-site spin density correlation function (see, e.g., Ref \onlinecite{TakahashiBook}). According to the fluctuation-dissipation theorem (FDT), SC can be obtained from the imaginary part of the dynamic magnetic susceptibility $\chi\left(\mathbf{q},\omega\right)$ by integrating over all wavevectors $\mathbf{q}$ and energies $\omega$. Experimentally $\chi\left(\mathbf{q},\omega\right)$ can be obtained from inelastic neutron scattering experiments, but such measurements can only probe energies up to 0.2-0.3 eV and typically access only limited parts of the Brillouion zone (BZ)\cite{Dai}. Correspondingly, only an approximate estimation of SC can be obtained experimentally. Nevertheless, the effective moment of about 1 $\mu_B$ was found in doped YMn$_2$ leading to a prediction of `giant spin fluctuations' in this material\cite{YMn2}. It is unclear, however, how these results would be modified if a larger energy range and a proper BZ integration would be included.

Theoretical treatments of itinerant SDF have also been restricted to both limited wavevectors and frequency range studies. If the traditional model spin fluctuation and/or quantum criticallity theories employed long wavelength and low frequency approximations (see e.g., Refs. \onlinecite{Hertz,Shimizu,Lonzarich,Solontsov,SolontsovBook,Solontsov2,Antropov}), the more recent electronic structure studies focused on pure intraatomic SDF using many-body perturbation theory\cite{Zein} or DMFT\cite{Kotliar} approaches. While applications of these two opposite limiting cases of SDF treatment have been successful in the description of many metallic systems, their essentially adjustable nature and uncontrollable approximations do not allow us to understand the relative roles of the different spatial or energy scales of SDF. Clearly, a comprehensive analysis of SDF in the entire BZ and for a wide energy range is needed. 

It is a goal of this paper to present such analysis using realistic electronic structure calculations. To focus on pure itinerant SDF, we consider a prototype system of 3\emph{d} paramagnets where the local moments are absent. We determine the strength and the character of such SDF as well as establish their spatial and energy scales which should be included for a proper description of ground state and thermodynamic properties of metals.

The dynamic spin susceptibility, $\chi(\mathbf{r},\mathbf{r}^{\prime},\mathbf{q},\omega)$, was evaluated using the linear response theory within the local density approximation\cite{Izuyama,Bergara}. We used an in-house electronic structure code based on the full-potential linear augmented plane waves method\cite{Kutepov}. Mixed product basis set\cite{PB} was adapted. Calculations were done at finite temperature using the Matsubara technique. First, the Kohn-Sham (or 'bare') susceptibility $\chi_0$ was evaluated in the real space and in the Matsubara time domain following the approach that was used for calculations of the polarizability in Ref. \onlinecite{Kutepov2}. Next, the Kohn-Sham susceptibility was transformed to the reciprocal space and the Matsubara frequency domain. The enhancement factor was then calculated and $\chi(\mathbf{r},\mathbf{r}^{\prime},\mathbf{q},\omega)$ was found. The real frequency axis results were obtained using the analytical continuation based on continued fraction expansion\cite{Vidberg}.

Paramagnetic SDF can be described in terms of the following spectral function

\begin{equation}
A(\mathbf{q},\omega)= -\frac{3}{\pi}\int d\mathbf{r}\int d\mathbf{r}^\prime\ \text{Im}\chi(\mathbf{r},\mathbf{r} ^{\prime},\mathbf{q},\omega) 
\label{A}
\end{equation} 

Here the factor of three corresponds to a number of polarizations and we integrated the $\mathbf{r}$ and $\mathbf{r}^{\prime}$ variables over the atomic sphere. Further we also define the density of on-site SDF $N(\omega)$ and the number of on-site SDF $n(\omega)$ as follows

\begin{equation}
n(\omega) =\int_{0}^{\omega }d\omega N(\omega)=\sum_{\mathbf{q}}^{BZ}\int_{0}^{\omega}d\omega A(\mathbf{q},\omega). 
\label{Nndefs}
\end{equation} 

Correspondingly, when the Kohn-Sham susceptibility is used in Eqs. (\ref{A}) and (\ref{Nndefs}), such functions are referred to as 'bare' quantities and we denote them by subscript $0$.  

An important measure of the strength of SDF is provided by the SC which is defined as the equal-time on-site spin density correlation function and can be obtained from $N(\omega)$ using the FDT

\begin{equation}
\left\langle\mathbf{s}^{2}\right\rangle =\int _{0}^{\infty }\coth{(\beta\omega/2)}N (\omega) d\omega.
\label{FDT}
\end{equation} 

The evaluation of the SC by a straightforward energy integration is problematic since the integrand in Eq. (\ref{FDT}) converge very slowly with energy. Therefore, we used a contour technique and performed the energy integration on the imaginary frequency axis\cite{Kutepov2} resulting in a much better convergence and an accurate SC.

We consider 3\emph{d} paramagnets (Sc, Ti, V, Cr, Cu, and Zn) at room temperature. For Cr a paramagetic potential was used. Experimental crystal structures were assumed with hcp for Sc, Ti, and Zn, fcc for Cu, and bcc for V and Cr. For cubic and hcp lattices we used 20$ \times $20$ \times $20 and 12$ \times $12$ \times $12 k-point mesh, respectively. For all materials, the local orbitals for the 3\emph{s} and 3\emph{p} semicore states were added to the basis function set. In the case of Cu and Zn the local orbitals for the 4\emph{d} states were included as well. For the  basis, the energy cutoff in the interstitial region was set to 2.5 Ry and the angular momentum cutoff inside the muffin-tin sphere was set to $L_{\text{max}} =4$. For the product basis functions we used the interstitial energy cutoff of 3.5 Ry and the muffin-tin angular momentum cutoff of $L_{\text{max}}^{\text{PB}}=4$. We used 158 nonuniformly distributed (see Ref. \onlinecite{Kutepov2} for details) mesh points on the imaginary Matsubara time axis. We ensured that the results are well converged with respect to the above parameters.

\begin{figure}[t!]
\includegraphics[width=1.0\hsize]{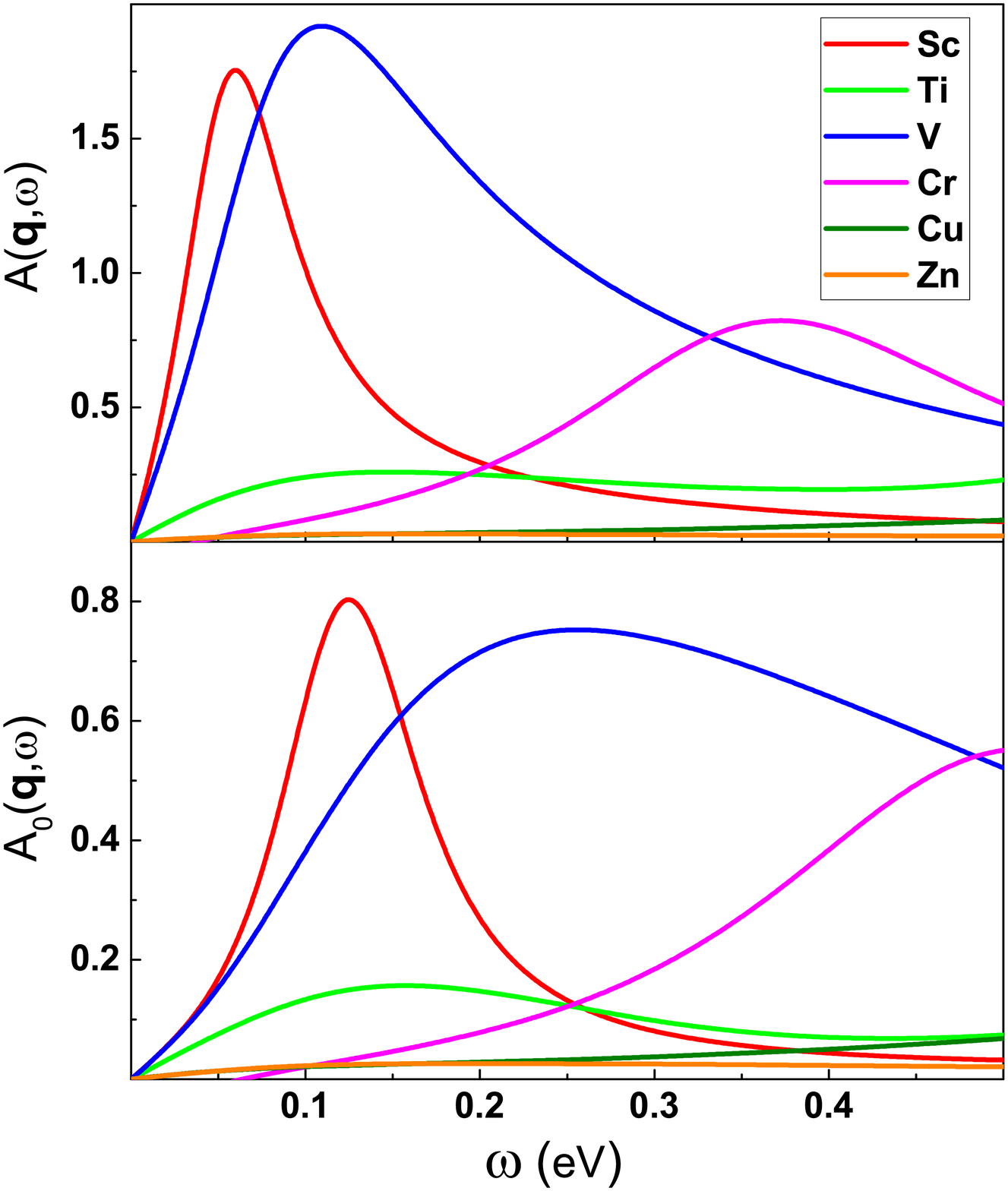}
\caption{Small wavevector SDF for different 3\emph{d} paramagnetic metals. (Top) Spectral function. (Bottom) 'Bare' spectral function. For cubic and hcp metals we used $\mathbf{q}=(0,0,0.1)2\pi/a$ and $\mathbf{q}=(0,0,0.05)2\pi/c$, respectively. The vertical axis units are $\hbar^2/\text{eV}$.}
\label{Paramagnons}
\end{figure}

Let us first consider SDF for small wavevectors. Fig. \ref{Paramagnons} shows the spectral function for a fixed low magnitude $\mathbf{q}$ as a function of the frequency within the neutron scattering energy range. For Sc, $A(\mathbf{q},\omega)$ has a broad peak at around 0.05 eV. A similar feature (although at higher energy) is observed for the 'bare' spectral function $A_{0}(\mathbf{q},\omega)$ suggesting that the peak predominantly originates from single particle Stoner excitations. More specifically, an analysis of the Sc band structure\cite{ScBandStructure} suggests that this feature corresponds to electronic transitions within the flat band that crosses the Fermi level near the H point in the BZ. The many body effects, however, still play an important role by significantly enhancing the peak and shifting it to lower energies. For V, $A(\mathbf{q},\omega)$ has a broad maximum at around 0.12 eV with the width of around 0.25 eV. $A_{0}(\mathbf{q},\omega)$ has a somewhat similar feature at higher energies indicating that, as in the case of Sc, the peak originates from the Kohn-Sham susceptibility but is enhanced and shifted to lower energies by electronic correlations. For other materials, the spectral functions in Fig.\ref{Paramagnons} are featureless. This indicates that the SDF for these wavevectors are also controlled by Stoner excitations.

\begin{figure}[t!]
\includegraphics[width=1.0\hsize]{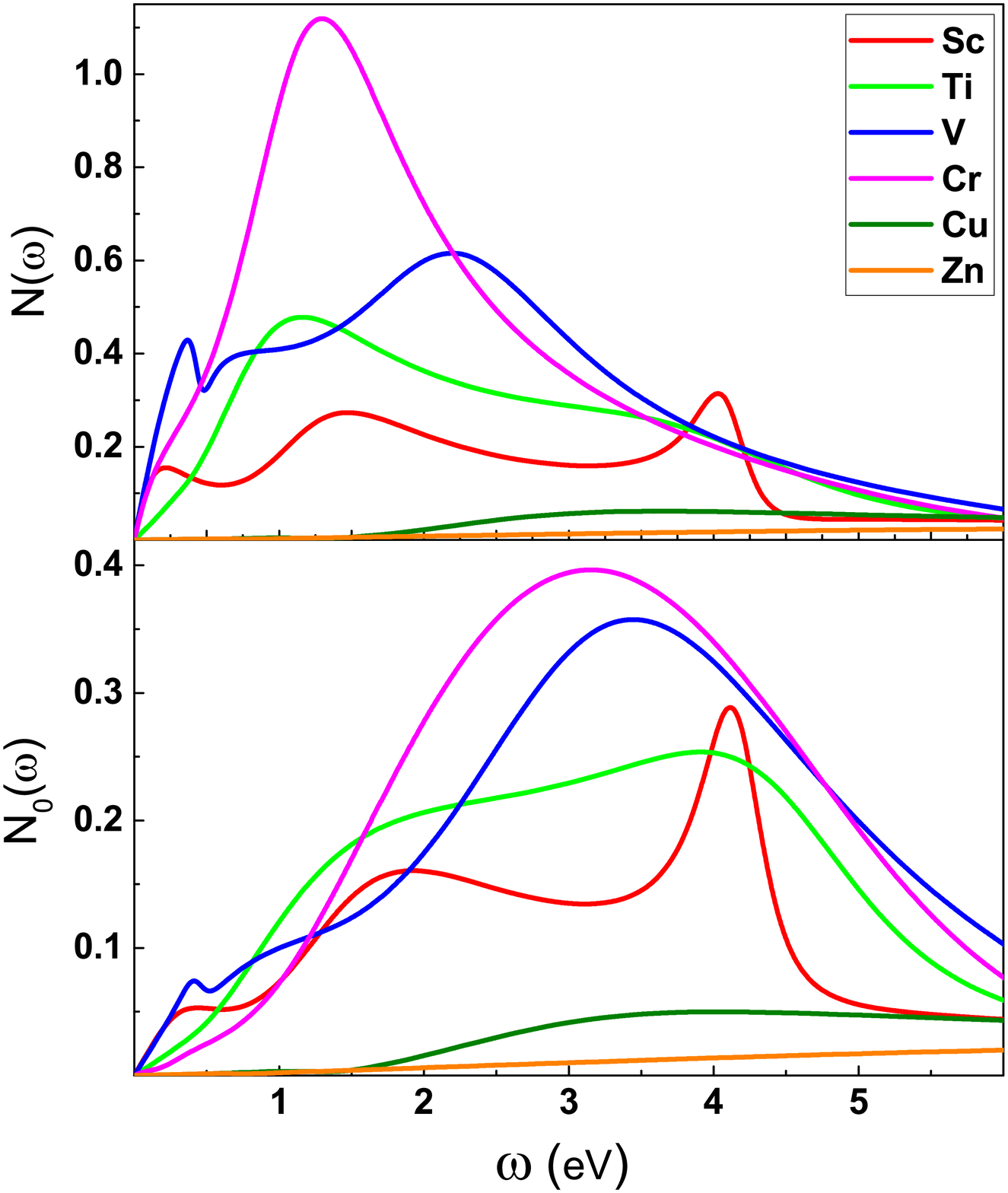}
\caption{On-site SDF spectrum for different 3\emph{d} paramagnetic metals. (Top) Density of SDF. (Bottom) 'Bare' density of SDF. The vertical axis units $\hbar^2/\text{eV}$.}
\label{Nsf}
\end{figure} 

Information about the SDF for other wavevectors can be obtained by analyzing the density of SDF (Fig. \ref{Nndefs}) which, according to Eq. (\ref{Nndefs}), include SDF from the entire BZ. Both $N(\omega)$ and $N_0(\omega)$ have a completely different shape from the corresponding spectral function. This indicates that the excitations with large wavevectors also play an important role. In fact, detailed analysis of our data revealed that SDF in all materials reside in the entire BZ. Therefore, restriction to excitations from only limited parts of the BZ (for instance the long wave approximation commonly used in spin fluctuations theories or DMFT single-site approximation) can lead to an inaccurate description of SDF. 

Except for Cu and Zn, the $N(\omega)$ curves can be viewed as superpositions of several peaks. Most of these peaks lie at relatively high energies and, therefore, one may expect that they originate from single-particle Stoner excitations. The spectrum of such excitations is described by $N_{0}(\omega)$. Evidently, $N_{0}(\omega)$ clearly shows a significant amount of single particle SDF. However, $N_{0}(\omega)$ is equal to $N(\omega)$ only for energies larger than the effective electronic bandwidth ($W_{el} \simeq$ 5 eV). For $\omega < W_{el}$ the two differ significantly. In particular, the electronic correlations significantly enhance the the single-particle spectrum and shift it to lower energies. Therefore, we conclude that in the energy range up to $W_{el}$ SDF in 3\emph{d} paramagnetic metals are a complicated mixture of mutually interacting single-particle and collective excitations. Consequently, realistic paramagnetic SDF cannot be treated as pure Bose excitations (as usually assumed\cite{Overhauser}).

\begin{figure}[t!]
\includegraphics[width=1.0\hsize]{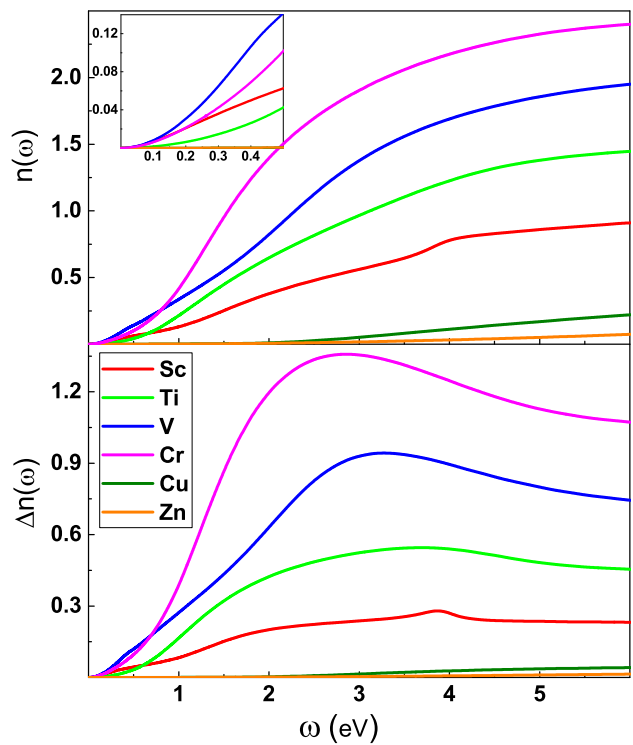}
\caption{(Top) Number of on-site SDF for different 3\emph{d} paramagnetic metals.  The inset shows the low-energy portion of the plot. For Cu and Zn the number of SDF is not seen in the inset since it is essentially zero at low energies. (Bottom) $\Delta n(\omega)=n(\omega)-n_0(\omega)$ as a function of energy. The vertical axis units are $\hbar^2$.}
\label{M2}
\end{figure} 

An important conclusion that follows from Fig. \ref{Nsf} is that SDF exist in a wide energy range extending up to several eV. This is even better illustrated in the top panel of Fig. \ref{M2} where the number of SDF, $n(\omega )$, (Eq. \ref{Nndefs}) is plotted as a function $\omega$. The main plot shows the full spectrum while the inset presents the low-energy range that is accessible to neutron scattering experiments.  As seen, the low-energy part (picosecond time scales) constitutes only a small fraction of the full SDF spectrum. Indeed, for early 3\emph{d} paramagnets (Sc, Ti, V, Cr) approximately above 0.5 eV, $n(\omega)$ starts to rapidly increase until the energy $W_{el}$ is reached. In this ultrafast energy range (femtosecond time scale), the majority of SDF reside. For $\omega > W_{el}$, only a slow increase of $n(\omega)$ is observed. This corresponds to SDF involving semicore and/or high-energy unoccupied states. These high-energy transitions are main contributions to SDF for Cu and Zn where the 3\emph{d} band is almost fully occupied. Therefore, for all 3\emph{d} paramagnets, the majority of SDF lie at energies much higher than those accessible from inelastic neutron scattering experiments requiring ultrafast experiments (high-energy spin resolved spectroscopies\cite{exp}) to probe the full SDF spectrum. We emphasize that for all considered materials, $n(\omega)$ is a continuous steadily increasing function of energy and it is not possible to rigorously introduce any energy cutoff when describing SDF in paramagnetic metals. Overall, these materials represent a slow-fast spin system with a strong interaction between time scales so adiabaticity criterion is not fulfilled. Thus, for instance,  with the temperature increase, more SDF are excited and contribute to the magnetic properties of the itinerant metal. This feature is in stark contrast with the traditional slow-fast system of magnetic insulator where time scales are well separated so that SDF for energies above spin wave spectrum do not exist and \emph{all} SDF are excited at corresponding temperatures.

Let us now consider SC  which is a major characteristic of the SDF and is related to the effective paramagnetic atomic moment ($m_{\text{eff}}$) by $m_{\text{eff}}=g\mu_B\sqrt{\left\langle\mathbf{s}^{2}\right\rangle}$. According to Fig. \ref{M2} (top), a dominant contribution to SC arises from SDF lying in the energy range from 0.5 eV up to $W_{el}$. In this regime, $\omega>>T=300 K$  and, thus, the SC originates mainly from spin zero-point motion SDF. However, even low energy SDF can create a significant SC which can be detected experimentally. For example, as seen from the inset in Fig. \ref{M2}, spin zero-point motion contribution to SC (given by $n(\omega)$) at the energy of 0.3 eV results in $m_{\text{eff}}$ of 0.3-0.6 $\mu_{B}$ for paramagnets with partially filled 3\emph{d} shell.

\begin{figure}[t!]
\includegraphics[width=1.0\hsize]{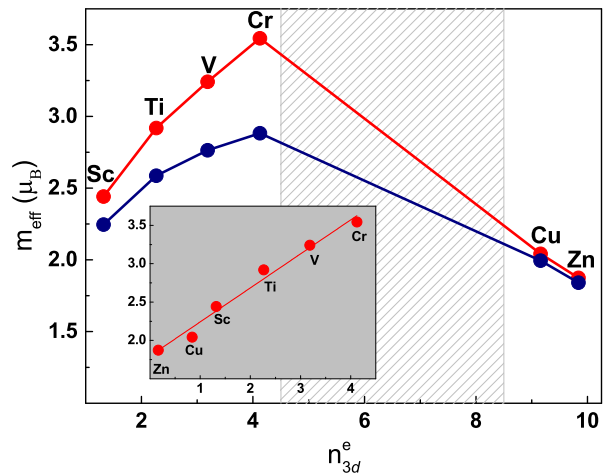}
\caption{Effective paramagnetic atomic moment ($m_{\text{eff}}$) as a function of the the number of 3\emph{d} electrons (red). Blue curve denote the 'bare' $m_{\text{eff}}$ evaluated using the Kohn-Sham susceptibility. The shaded area indicates the 3\emph{d} electron range corresponding to the magnetically ordered systems. The inset shows $m_{\text{eff}}$ as a function of $n_{3d}=\text{min}\left(n^{e}_{3d},n^{h}_{3d}\right)$ where $n^{e}_{3d}$ and $n^{h}_{3d}$ is the number of 3\emph{d} electrons and holes, respectively. The line in the inset is the linear fit of the data.}
\label{SPcurve}
\end{figure}

Fig. \ref{SPcurve} shows $m_{\text{eff}}$ (evaluated from the entire SDF spectrum) for all 3\emph{d} paramagnets.  As seen, for all considered materials, $m_{\text{eff}}$ has a value of several $\mu_{B}$.  However, the 'bare' $m_{\text{eff}}$ that originate purely from single particle SDF is also appreciable. In particular, for Cu and Zn, it is essentially equal to $m_{\text{eff}}$. This implies a small contribution of SDF to the correlations. For other 3\emph{d} paramagnets, however, the difference between $m_{\text{eff}}$ and the 'bare' $m_{\text{eff}}$ is appreciable (1$\mu_{B}$ for Cr). Therefore, a significant effect of SDF on all ground state and thermodynamic properties is expected in these materials. Indeed, the measure of the SDF correlation energy is provided by $I\Delta n(\infty)/2$ with $I$ being an interaction parameter and $\Delta n(\omega)=n(\omega)-n_0(\omega)$ (see, for instance, recent review in Ref. \onlinecite{RPA}). As seen in the bottom panel of Fig. \ref{M2}, while $\Delta n(\omega)$ is negligible for both Cu and Zn, it is large for other metals and at the energy of the order $W_{el}$ it practically reaches its  total value. Therefore, for paramagnets with partially filled 3\emph{d} shell, SDF corresponding to this $\Delta n(\omega)$ should be explicitly included in electronic structure calculations and all the energies up to $W_{el}$ should be treated on equal footing. We emphasize that for all considered materials the local moment is zero and, therefore, these SDF are purely itinerant and cannot be described by the Heisenberg model. Naturally, one can expect that in ferromagnetic metals such itinerant SDF also play a significant role and, correspondingly, should be taken into account simultaneously with the Heisenberg local moment fluctuations.

Let us analyze how SDF depend on the 3\emph{d} band population. As seen from Fig. \ref{M2}, the low energy part of $n(\omega)$ doesn't show any clear dependence on the 3\emph{d} band filling since it is mainly determined by the detailed structure of the Fermi surface and, thus, it is a very material specific part of the spectrum. However, for large energies, the intensity of SDF increases with a number of unpaired 3\emph{d} electrons. Correspondingly, the dependence of $m_{\text{eff}}$ on the 3\emph{d} electron number is reminiscent of the Slater Pauling curve for ferromagnetic metals (Fig. \ref{SPcurve}). Below the half-filling, $m_{\text{eff}}$ increases with the 3\emph{d} electron number. Above the half-filling, however, an opposite trend is observed. Two mechanisms that are similar in strength but differ in origin are responsible for such behavior. First, as indicated by the blue curve in Fig. \ref{SPcurve}, the number of Stoner single particle excitations within the 3\emph{d} band is maximized at half-filling. Second, the many body enhancement due to electronic correlations is also the strongest for a half-filled band. 

In the inset of Fig. \ref{SPcurve}, we show $m_{\text{eff}}$ as a function of the effective number 3\emph{d} carriers $n_{3d}=\text{min}\left(n_{3d}^{e},n_{3d}^{h}\right)$ where $n_{3d}^{e}$ and $n_{3d}^{h}$ is the number of 3\emph{d} electrons and holes, respectively. Interestingly, we find that $m_{\text{eff}}$ shows approximately a linear dependence on $n_{3d}$ according to the following empirical formula 

\begin{equation}
m_{\text{eff}} \approx 0.4 n_{3d}  +1.8.
\end{equation} 

The above indicates that every 3\emph{d} electron or hole contributes approximately the moment of 0.4$\mu_{B}$ to $m_{\text{eff}}$. The nonzero intercept corresponds to $m_{\text{eff}}$ for a completely filled or completely empty 3\emph{d} band. It originates from electronic transitions involving semicore levels and high-energy unoccupied states.

In summary, we investigated the structure of SDF for 3\emph{d} paramagnetic metals. A strong SC corresponding to the effective magnetic fluctuating moment of the order of the several Bohr magnetons was obtained. We found that SDF in these systems have a mixed nature and consist of interacting single particle Stoner and collective excitations. This indicates that no well defined quantum statistics can be assigned to these excitations. These SDF are rather strong and spread continuously over the entire BZ as well as the wide energy range. Therefore, such SDF can be accessed only by very different energy scale measurements ranging from neutron to ultrafast experiments. From a point of view of theory, no well defined wavevector and frequency cutoffs (as often assumed) can be reliably introduced in such materials. In particular, for the correct description of spin fluctuations induced phenomena in metals all such SDF should be included on equal footing without usage of any long wave length or atomistic approximations. 

\section*{Acknowledgments}
This work was supported by the Critical Materials Institute, an Energy Innovation Hub funded by the U.S. Department of Energy (DOE). V. P. acknowledges the support from the Office of Basic Energy Science, Division of Materials Science and Engineering. The research was performed at Ames Laboratory, which is operated for the U.S. DOE by Iowa State University under contract \# DE-AC02-07CH11358.

\end{document}